\documentclass[fp]{jpsj3}
\usepackage{txfonts}
\usepackage{bm}
\usepackage{amsmath}
\usepackage{amsfonts}
\usepackage{color}

\title{Orbital Order in Two-Orbital Hubbard Model}

\author{Kojiro Honkawa$^1$ and Seiichiro Onari$^{1,2}$\thanks{onari@okayama-u.ac.jp}}
\inst{$^1$Department of Physics, Okayama University, Okayama 700-8530,
Japan \\
$^2$Research Institute for Interdisciplinary Science, Okayama University, Okayama 700-8530, Japan}

\abst{In strongly correlated multiorbital systems, various ordered phases
appear. In particular, the orbital order in iron-based
superconductors attracts much attention since it is considered to be the origin of
the nematic state. To clarify the essential conditions for realizing orbital orders,
we study the simple two-orbital ($d_{xz}$, $d_{yz}$) Hubbard model.
We find that the orbital order, which corresponds to the nematic order, appears
due to the vertex corrections even in the two-orbital model. Thus, the $d_{xy}$ orbital is not essential to realize
the nematic orbital order. The
obtained orbital order
is determined by the orbital dependence and the topology of Fermi surfaces. We also find that another
type of orbital order, which is rotated $45^\circ$, appears in a heavily hole-doped case.
}

\begin{document}
\maketitle

\section{Introduction}
In the phase diagram of iron-based superconductors, the electronic nematic state with $C_2$
symmetry appears
universally under the structural phase transition temperature $T_{\rm
S}$, and the superconducting state emerges next to the nematic state \cite{phase1,phase2}.
The origin of the nematic state has
stimulated much attention since the mechanism of superconductivity
would be strongly related to the mechanism of the nematic state. 
To explain the nematic order and fluctuations, both the
spin-nematic scenario\cite{Fernandes,Fernandes2} and the orbital-order scenario\cite{Onari-SCVC,Onari-Hdoped,Kontani-Onari,Kontani-quad,Kruger,PP,WKu} have been
proposed. In the former scenario, the origin of the nematic state is the
spin-nematic order. The spin-fluctuation-induced spin-nematic order
appears slightly above the magnetic transition temperature when the magnetic
frustration is strong. However, the nematic state in FeSe with small
spin fluctuations cannot be explained by the spin-nematic scenario.
In the latter scenario, the origin is the ferro-orbital
 order ($n_{xz}\ne n_{yz}$), which is caused by the
 vertex correction (VC) of the Coulomb interaction
 \cite{Onari-SCVC,Onari-Hdoped}. 
By applying the self-consistent VC (SC-VC) method to the realistic five
Fe $d$ orbital model based on first-principles calculation, the
nematic state in both
LaFeAsO and FeSe has been
naturally reproduced\cite{Onari-Hdoped,FeSe-Yamakawa2}.

Recently, the remarkable
$\bm{k}$ dependence of the orbital polarization, which cannot be
explained using the mean-field theory, has been observed in FeSe\cite{FeSe-ARPES62}.
The $\bm{k}$ dependence of the orbital polarization is also
reproduced in the orbital-order scenario\cite{Onari-FeSe}.
In addition, the $s$-wave state 
without sign reversal ($s_{++}$-wave state) mediated by the orbital
fluctuations has also been
proposed in iron-based superconductors \cite{Onari-SCVC,Onari-Hdoped,Kontani-Onari,Kontani-quad,Saito,Saito2}. 
The SC-VC theory had also succeeded in explaining other multiorbital
systems such as the nematic CDW in cuprates \cite{cuprate1,cuprate2},
the nematic state in Sr$_3$Ru$_2$O$_7$ \cite{Sr327}, and the spin-triplet
superconductivity in Sr$_2$RuO$_4$ \cite{Sr214-1,Sr214-2}.
%By taking account of the Coulomb interaction 
%and the nesting of the Fermi surfaces (FSs) a fully-gapped sign-reversing $s$-wave state ($s_\pm$-wave state) had been proposed 
%based on the spin fluctuation theories
% \cite{Kuroki,Mazin,Tesanovic,Hirschfeld,Chubukov}.
%However, $s_\pm$-wave superconducting (SC) state is fragile 
%against impurities \cite{Onari-impurity}, 
%inconsistently with experiments \cite{Sato-imp}.

In our previous work based on the SC-VC theory in realistic five-orbital models of iron-based superconductors, we have found that the nematic
orbital order $n_{xz}\ne n_{yz}$ is caused by the VC including
$d_{xz}$, $d_{yz}$, and $d_{xy}$ orbitals \cite{Onari-Hdoped,FeSe-Yamakawa2}. In
the five-orbital models, Fermi surfaces (FSs) are composed of $d_{xz}$, $d_{yz}$ and
$d_{xy}$ orbitals. Thus, these orbitals are expected to be important in realizing the
nematic orbital order.
However, the necessary conditions for the orbital order are yet unclear.
In order to clarify the essence of the orbital order, we study the
simple two-orbital ($d_{xz}$, $d_{yz}$) model systematically by the SC-VC method.

We found that the VC for the orbital susceptibility is also enhanced in the simple two-orbital model,
and that the nematic orbital order appears similarly to the case of the realistic
five-orbital models. Thus, the $d_{xy}$ orbital is not essential to realize
the nematic orbital order. We also found that another type of orbital order,
which corresponds to the $45^\circ$-rotated orbital order, appears when the
 holes are heavily doped. The obtained orbital order is determined by the orbital dependence and the
topology of the FSs.

\section{Formulation}
We study a simple two-orbital ($d_{xz}$, $d_{yz}$) Hubbard model in a square
lattice.
The Hamiltonian is given as
\begin{eqnarray}
H=\sum_{\bm{k}}\sum_{l,m=1,2}\sum_{\sigma=\uparrow,\downarrow}\xi^{lm}_{\bm{k}}c^\dagger_{\bm{k},l,\sigma}c_{\bm{k},m,\sigma}+H_{\rm
 int},
\end{eqnarray}
where $l,m=1,2$ represents the orbital; $1=d_{xz}$ and $2=d_{yz}$.
In the tight-binding model, we employ the dispersion relation\cite{Raghu} 
$\xi^{11}_{\bm{k}}=-2t_1\cos(k_x)-2t_2\cos(k_y)-4t_3\cos(k_x)\cos(k_y)$,
$\xi^{22}_{\bm{k}}=-2t_2\cos(k_x)-2t_1\cos(k_y)-4t_3\cos(k_x)\cos(k_y)$,
and $\xi^{12}_{\bm{k}}=\xi^{21}_{\bm{k}}=-4t_4\sin(k_x)\sin(k_y)$.
We control the band filling $n$ systematically, where $n=1$ corresponds to
the half-filling.
$H_{\rm int}$ is the multiorbital Coulomb interaction including the intra (inter)
orbital interaction $U(U')$ and the exchange interaction $J$. Hereafter, we
use the relation $U=U'+2J$ from rotational invariance.
The irreducible susceptibility in the orbital representation without the VC is given as
$\chi_{ll',mm'}^0(q)=-\frac{T}{N}\sum_k{G}_{lm}(k+q){G}_{m'l'}(k)$,
where $G$ is the Green's function, and $q$ and $k$ are denoted as
$q=(\bm{q},\omega_n=2n\pi T)$ and $k=(\bm{k},\epsilon_n=(2n+1)\pi T)$,
respectively. 
{\color{black}We take $N=64\times64$ $\bm{k}$ meshes and 256 Matsubara
frequencies.}
Using the RPA, the matrix of the spin (orbital) susceptibility is given
as
{\color{black}
\begin{eqnarray}
\hat{\chi}^{s(c)}(q)=\hat{\chi}^0(q)\left[1-\hat{\Gamma}^{s(c)}\hat{\chi}^0(q)\right]^{-1},
\end{eqnarray}
}
where $\hat{\Gamma}^{s(c)}$ denotes the bare Coulomb interaction for spin (charge) channel given by 
\begin{eqnarray}
\Gamma_{l_{1}l_{2},l_{3}l_{4}}^{s(c)} = \begin{cases}
U(-U), & l_1=l_2=l_3=l_4 \\
U'(U'-2J) , & l_1=l_3 \neq l_2=l_4 \\
J(-2U'+J), & l_1=l_2 \neq l_3=l_4 \\
J(-J) , & l_1=l_4 \neq l_2=l_3
\end{cases}.
\label{eqn:Gamma-s}
\end{eqnarray}

By introducing the VC $\hat{X}^{c}$ for the orbital channel,
the matrix of the orbital susceptibility is given by
{\color{black}
\begin{eqnarray}
\hat{\chi}^{c}(q)=\left[\hat{\chi}^0(q)+\hat{X}^{c}(q)\right]\left[1-\hat{\Gamma}^{c}\left\{\hat{\chi}^0(q)+\hat{X}^{c}(q)\right\}\right]^{-1}\label{chi}.
\end{eqnarray}
}
In this study, we neglect the VC $\hat{X}^{s}$ for the spin channel
for simplicity
since $\hat{X}^{s}$ is smaller than $\hat{X}^{c}$, as reported previously\cite{Onari-SCVC}.
The spin Stoner factor $\alpha_{s}$ is given by
 the maximum eigenvalue of $\hat{\Gamma}^{s}\hat{\chi}^0(\bm{q},0)$,
 while the charge Stoner factor $\alpha_{c}$ is given by the maximum
 eigenvalue of 
 $\hat{\Gamma}^{c}\left[\hat{\chi}^0(\bm{q},0)+\hat{X}^{c}(\bm{q},0)\right]$. 
{\color{black} The ordered state is realized when $\alpha_{s}=1$ or
$\alpha_{c}=1$.}

The RPA is recovered by putting $\hat{X}^{c}=0$.
It is clear that $\hat{\chi}^{c}(q)$ is enhanced over the value
in the RPA when $\hat{X}^{c}(q)$ is large.
In the SC-VC method, $\hat{\chi}^{c,s}$ and $\hat{X}^{c}$ are obtained self-consistently.
The VC is given by the Maki-Thompson (MT) and Aslamazov-Larkin (AL)
terms, which are the first- and
second-order terms, respectively, with respect to $\hat{\chi}^{s(c)}$. 
The MT term for the orbital susceptibility $X_{ll',mm'}^{{\rm
MT},c}(q)$ is given by 
{\color{black}
\begin{eqnarray}
X_{ll',mm'}^{{\rm
 MT},c}(q)\!\!\!\!&=&\!\!\!\!-\left(\frac{T}{N}\right)^2\sum_{k,k'}\sum_{a,b,c,d}G_{la}(k+q)G_{cl'}(k)G_{bm}(k'+q)G_{m'd}(k')\nonumber\\
&&\!\!\!\!\left[\frac{1}{2}V^c_{ab,cd}(k-k')+\frac{3}{2}V^s_{ab,cd}(k-k')-\frac{1}{2}\Gamma^c_{ab,cd}-\frac{3}{2}\Gamma^s_{ab,cd}\right], \label{MT-c}
\end{eqnarray}
}
where ${\hat V}^{s,c}(q)\equiv{\hat \Gamma}^{s,c}
+ {\hat\Gamma}^{s,c}{\hat\chi}^{s,c}(q){\hat\Gamma}^{s,c} $.

The AL term for the orbital susceptibility $X_{ll',mm'}^{{\rm AL},c}(q)$
is given by
{\color{black}
\begin{eqnarray}
X_{ll',mm'}^{{\rm AL},c}(q)\!\!\!\!&=&\!\!\!\!\frac{T}{N}\sum_{q'}\sum_{a\sim h}
\Lambda_{ll',ab,ef}(q;q')\left[  \frac{3}{2}{V}_{ab,cd}^s(q+q'){V}_{ef,gh}^s(-q')\right.
\nonumber \\
& &\!\!\!\!\left.\!\!\!\!+\frac{1}{2}{V}_{ab,cd}^c(q+q'){V}_{ef,gh}^c(-q') \right]
\Lambda_{mm',cd,gh}'(q;q') ,
 \label{AL-c}
\end{eqnarray}
}
where the three-point vertex ${\hat \Lambda}(q;q')$ is given as
{\color{black}
\begin{eqnarray}
\Lambda_{ll',ab,ef}(q;q')=-\frac{T}{N}\sum_{k}G_{la}(k+q)G_{fl'}(k)G_{be}(k-q'),
\end{eqnarray}
}
 and 
$\Lambda_{mm',cd,gh}'(q;q')\equiv
\Lambda_{ch,mg,dm'}(q;q')+\Lambda_{gd,mc,hm'}(q;-q-q')$.
{\color{black} When we calculate the total VC, we subtract the
double-counting $U^2$-terms in Eqs. (\ref{MT-c}) and (\ref{AL-c})\cite{Onari-Springer}.}
By calculating the above equations self-consistently, we find that the MT
 {\color{black}and $U^2$-terms} are much smaller than the AL terms.

%Here, we explain the reason why the orbital fluctuations is increased by the AL term.
%Since the AF spin fluctuations $\hat{\chi}^s(k)$ develop at the
%nesting vector $\bm{k}=\bm{Q}$, the first term of Eq.(\ref{Xc}) 
%$\sum_k\hat{\chi}^s(q+k)\hat{\chi}^s(-k)$ (two-magnon process) enhances
%$\hat{X}^{{\rm AL},c}(\bm{q})$  at $\bm{q}={\bm 0}$.

Here, we introduce the diagonal charge quadrupole 
susceptibilities as
\begin{eqnarray}
\chi_{\gamma}^{Q}(q)&=&
\sum_{ll'}\sum_{mm'}O_\gamma^{l,l'}\chi_{ll',mm'}^c(q)O_\gamma^{m,m'},
 \label{eqn:chiQ} 
\end{eqnarray}
where $O_\gamma^{l,m}\equiv\langle l|\hat{O}_\gamma|m\rangle$ is the
matrix element of the $\gamma$-type quadrupole operator. In the present
two-orbital system, $O^{1,1}_{x^2-y^2}=-O^{2,2}_{x^2-y^2}=1$, 
$O^{1,2}_{xy}=O^{2,1}_{xy}=1$, and
$O^{1,1}_{3z^2-r^2}=O^{2,2}_{3z^2-r^2}=1/2$,  while other matrix
elements are zero.
The nematic state with $n_{xz}\ne n_{yz}$ is explained by the divergent
of $\chi^Q_{x^2-y^2}(\bm{q}=\bm{0},0)$\cite{Onari-SCVC,Onari-Hdoped}. {\color{black}In
the two-orbital system without the $d_{xy}$ orbital, $\chi^Q_{3z^2-r^2}$
is identical to the charge susceptibility. Thus, the enhancement of
$\chi^Q_{3z^2-r^2}$ driven by the $d_{xy}$ orbital in the
five-orbital model\cite{Onari-Hdoped} is not realized in the two-orbital system.

 Hereafter, we fix the temperature $T=0.05$ and $J/U=0.05$. 
Although $J/U=0.05$ is smaller than the value obtained by
the first-principles calculation\cite{Miyake}, we note that the obtained
orbital fluctuations without the self-energy for a small $J/U\sim0.05$
are similar to those including the self-energy for a large $J/U\sim0.14$ in the five-orbital system\cite{Onari-Hdoped}. 
The VC for the orbital fluctuations is underestimated in the system without the self-energy
since $U$ is underestimated.}
\section{Results and Discussion}
\begin{figure}
\includegraphics[width=\linewidth]{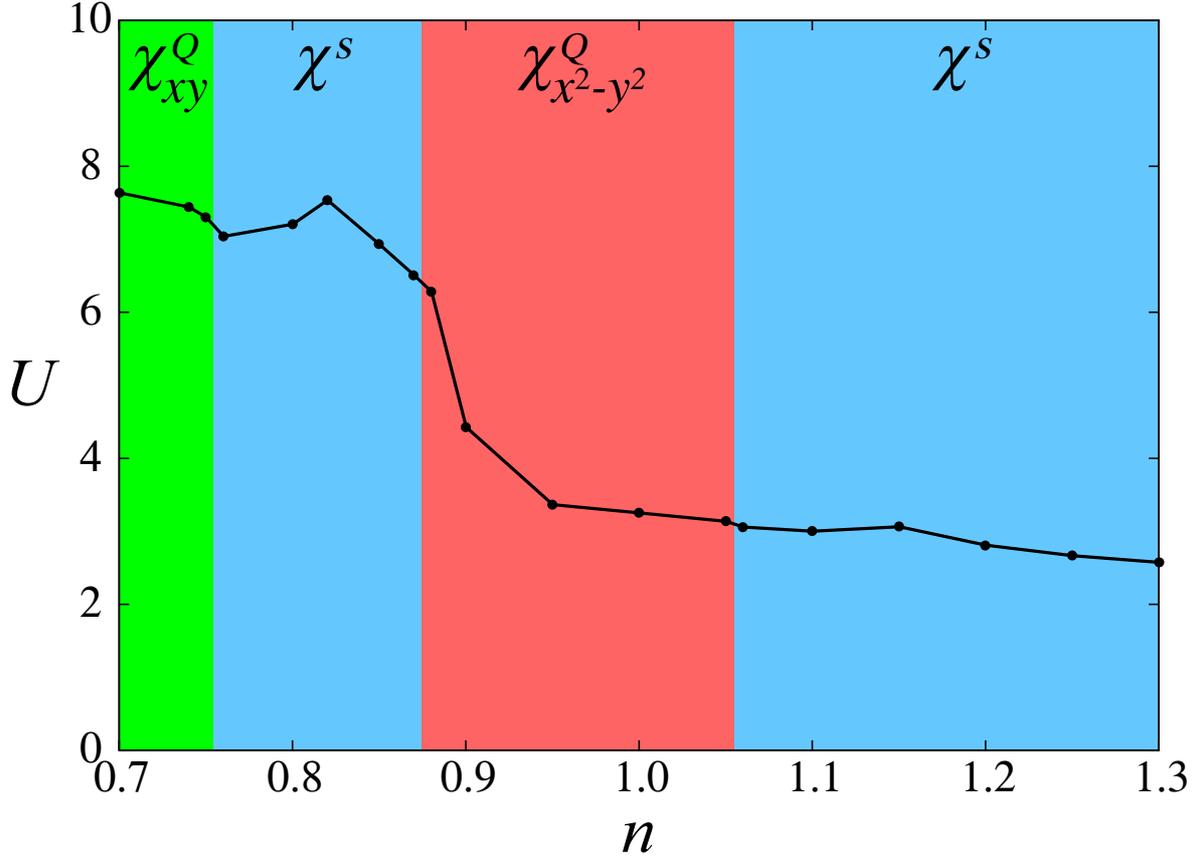}
\caption{(Color online) $U$ as a function of band filling $n$, where $\alpha_{s(c)}=0.97$ is satisfied for each $n$. In the blue region,
 the spin fluctuations are dominant. In the red (green) region, the
 $x^2-y^2$-type ($xy$-type) charge quadrupole fluctuations are dominant.
}
\label{phase-case1}
\end{figure}

{\color{black}We employ the two-orbital tight-binding model of iron-based
superconductors with $t_1=-1$, $t_2=1.39$, and
$t_3=t_4=-0.85$, which is similar to the model given in
Ref. \citen{Raghu}.} 
{\color{black} First, we calculate the susceptibilities near the ordered state by the SC-VC method.}
Figure \ref{phase-case1} shows $U$ as a function of $n$, where
$\alpha_s=0.97$ or $\alpha_c=0.97$ is satisfied for each $n$. 
In the blue region, $\alpha_s=0.97$ is satisfied and spin fluctuations
are dominant. 
In the red (green) region, $\alpha_c=0.97$ is satisfied and $x^2-y^2$-type ($xy$-type) charge quadrupole
fluctuations are dominant. In this model, the $3z^2-r^2$-type charge
quadrupole susceptibility, {\color{black}which is identical to the charge
susceptibility,} is
much smaller than other susceptibilities. {\color{black}In
Fig. \ref{phase-case1}, the small $U$ indicates that the corresponding ordered
state is easily realized.
Although we cannot access $\alpha_{s(c)}>0.98$ in order
to maintain the  calculation accuracy, we confirmed that the dominant fluctuations
are basically independent of $\alpha_{s(c)}$ for $0.98\ge \alpha_{s(c)}\ge 0.95$. Only the boundary of the
regions slightly depends on $\alpha_{s(c)}$.}
For $n=1.0$, the $x^2-y^2$-type charge quadrupole
fluctuation with the peak at $\bm{q}=(0,0)$ is dominant over the spin
fluctuations and the $xy$-type
charge quadrupole fluctuations. This
result is consistent with the phase diagram of iron-based superconductors,
where the nematic $x^2-y^2$-type charge quadrupole (orbital) order
appears at higher temperatures than at which the spin order appears.

\begin{figure}
\includegraphics[width=\linewidth]{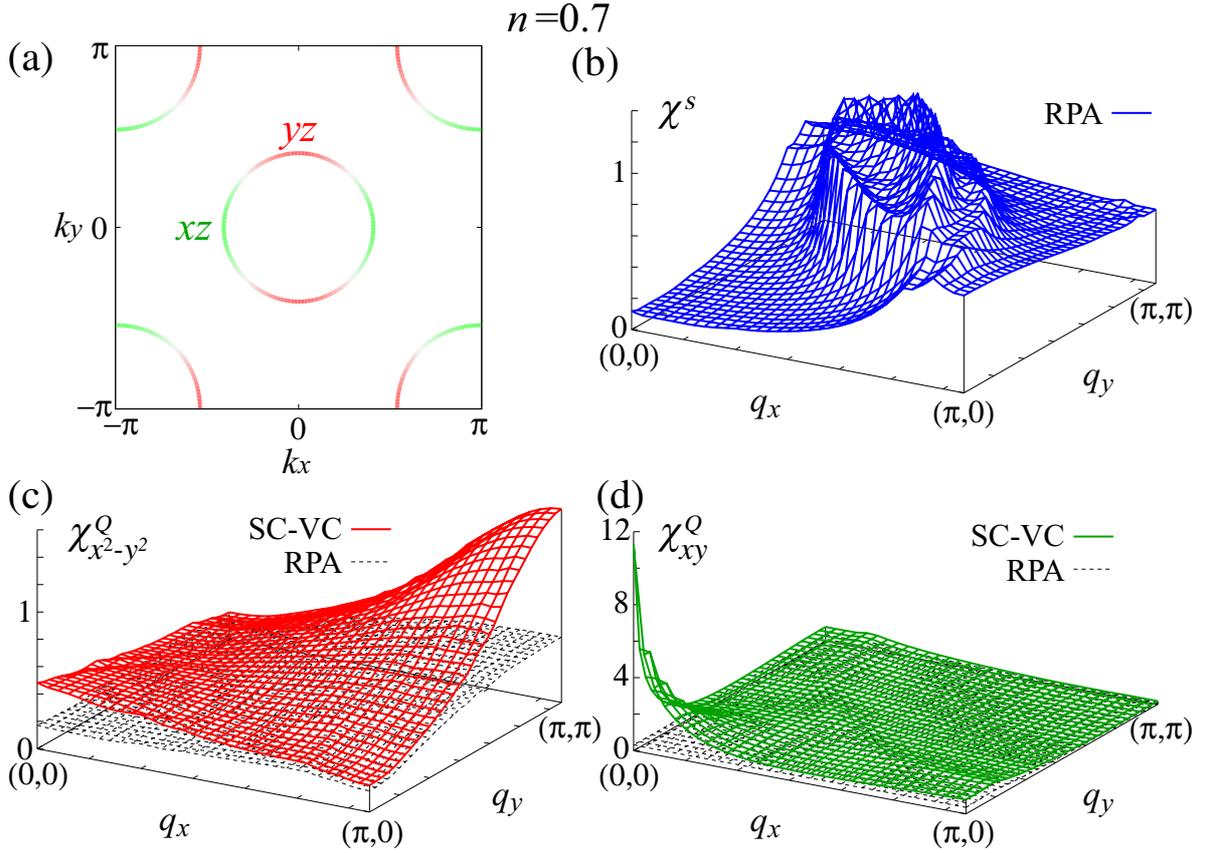}
\caption{(Color online) (a)FSs composed of orbital 1 (green line) and orbital
 2 (red line), {\color{black}(b)$\chi^s(\bm{q},0)$ in the RPA,
 (c)$\chi^Q_{x^2-y^2}(\bm{q},0)$ in the SC-VC method (solid red line) and the
 RPA (black dotted line), and (d)$\chi^Q_{xy}(\bm{q},0)$ in the SC-VC
 method (solid green line) and the
 RPA (black dotted line) for $n=0.7$.}
}
\label{n07-case1}
\end{figure}

In Fig. \ref{n10-case1}(a), we show FSs composed of orbital 1 (green
line) and orbital 2 (red line) for $n=1.0$. The FSs around $\bm{k}=(0,0)$ and
$\bm{k}=(\pi,\pi)$ are the hole FSs, while the FSs around $\bm{k}=(\pi,0)$ and $\bm{k}=(0,\pi)$ are
the electron FSs.
These FSs are consistent with the FSs of
iron-based superconductors except for the FS around $\bm{k}=(\pi,\pi)$, which is
composed of $d_{xy}$ orbital in iron-based superconductors. Figures
\ref{n10-case1}(b)--\ref{n10-case1}(d) show the $\bm{q}$ dependences of
$\chi^s(\bm{q},0)$, $\chi^Q_{x^2-y^2}(\bm{q},0)$, and $\chi^Q_{xy}(\bm{q},0)$, respectively.
$\chi^s(\bm{q},0)$ has a peak at $\bm{q}=(\pi,0),(0,\pi)$ due to the
strong intraorbital nesting between the hole FSs and the electron FSs. 
For $n=1.0$, $\chi^Q_{x^2-y^2}$ with peak at $\bm{q}=(0,0)$ is dominant over $\chi^s$ and $\chi^Q_{xy}$, which is
consistent with the nematic order (fluctuation) in low-doped iron-based superconductors.
$\chi^Q_{x^2-y^2}(\bm{0},0)$ is enhanced by the AL terms
$X^{{\rm AL},c}_{11,11}(\bm{0},0),X^{{\rm AL},c}_{22,22}(\bm{0},0)$ in
Eq. (\ref{AL-c}). $X^{{\rm AL},c}_{11,11(22,22)}(\bm{0},0)$ is mainly enlarged by the
intraorbital spin fluctuation terms {\color{black}
${V}_{11,11(22,22)}^s[\bm{Q_1}(\bm{Q_2}),0]{V}_{11,11(22,22)}^s[-\bm{Q_1}(-\bm{Q_2}),0]$, where
$\bm{Q_1}=(0,\pi)$ and $\bm{Q_2}=(\pi,0)$ are the nesting vectors for
orbitals 1 and 2, respectively. The enhancement of
$\chi^Q_{xy}(\bm{q},0)$ around $\bm{Q_3}=(\pi,\pi)$ is caused by the
AL terms $X^{{\rm AL},c}_{12,12}(\bm{Q_3},0),X^{{\rm
AL},c}_{21,21}(\bm{Q_3},0)$.
$X^{{\rm AL},c}_{12,12(21,21)}(\bm{Q_3},0)$ is mainly enlarged by the
terms
${V}_{11,11(22,22)}^s[\bm{Q_1}(\bm{Q_2}),0]{V}_{12,12(21,21)}^s[\bm{Q_2}(\bm{Q_1}),0]$
and ${V}_{11,11(22,22)}^c(\bm{0},0){V}_{12,12(21,21)}^c(\bm{Q_3},0)$ in
Eq. (\ref{AL-c}).

\begin{figure}
\includegraphics[width=\linewidth]{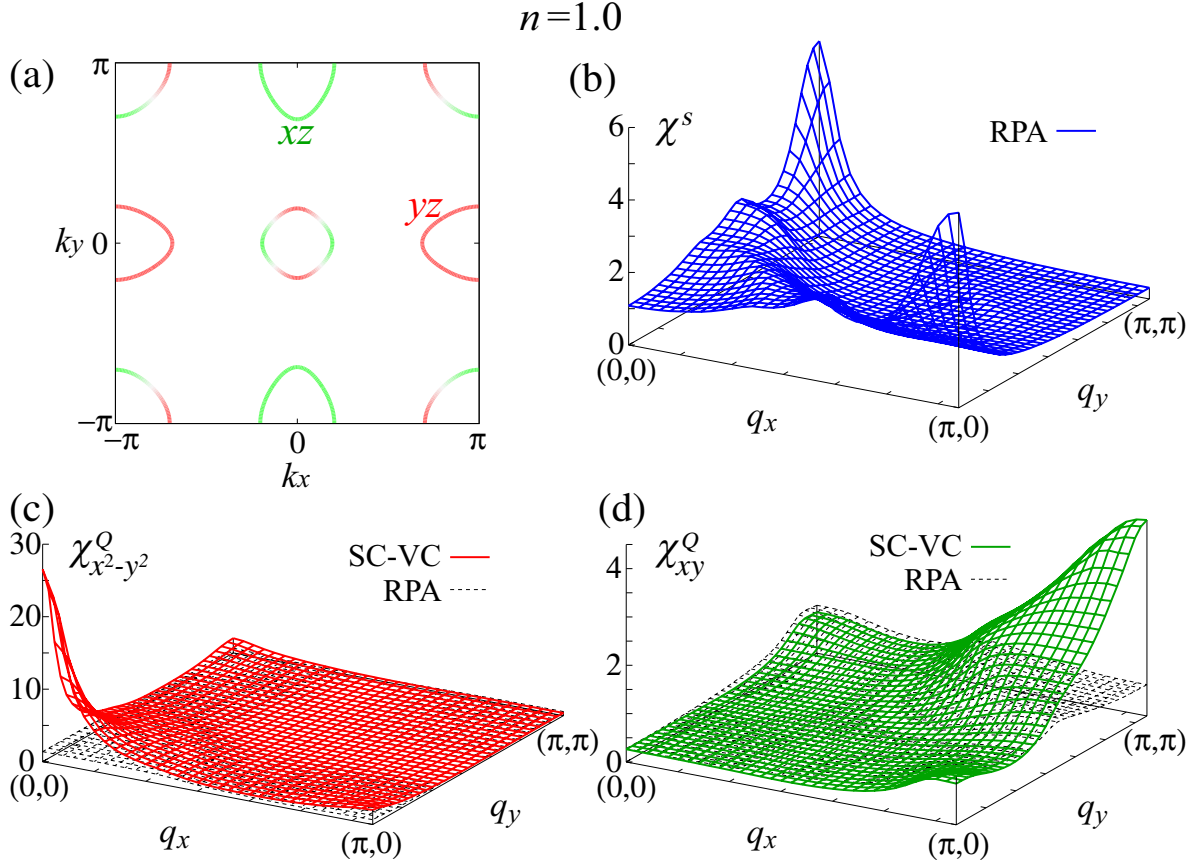}
\caption{(Color online) (a)FS composed of orbital 1 (green line) and orbital
 2 (red line), {\color{black}(b)$\chi^s(\bm{q},0)$ in the RPA,
 (c)$\chi^Q_{x^2-y^2}(\bm{q},0)$ in the SC-VC method (solid red line) and the
 RPA (black dotted line), and (d)$\chi^Q_{xy}(\bm{q},0)$ in the SC-VC
 method (solid green line) and the
 RPA (black dotted line) for $n=1.0$.}
}
\label{n10-case1}
\end{figure}

Figure \ref{n12-case4} shows the results for $n=1.2$.
$\chi^s$ has a peak around $\bm{q}=(\pi,\pi/5),(\pi/5,\pi)$ due to the
intraorbital nesting between the hole FSs and the electron FSs. This
incommensurate spin fluctuation originates from the large
difference in diameter between the hole FSs and the electron-FSs.
In this filling, $\chi^s$ is dominant over $\chi^Q_{x^2-y^2}$ and
$\chi^Q_{xy}$. The enhancement of $\chi^Q_{x^2-y^2}$ and
$\chi^Q_{xy}$ by the VC is small.

\begin{figure}
\includegraphics[width=\linewidth]{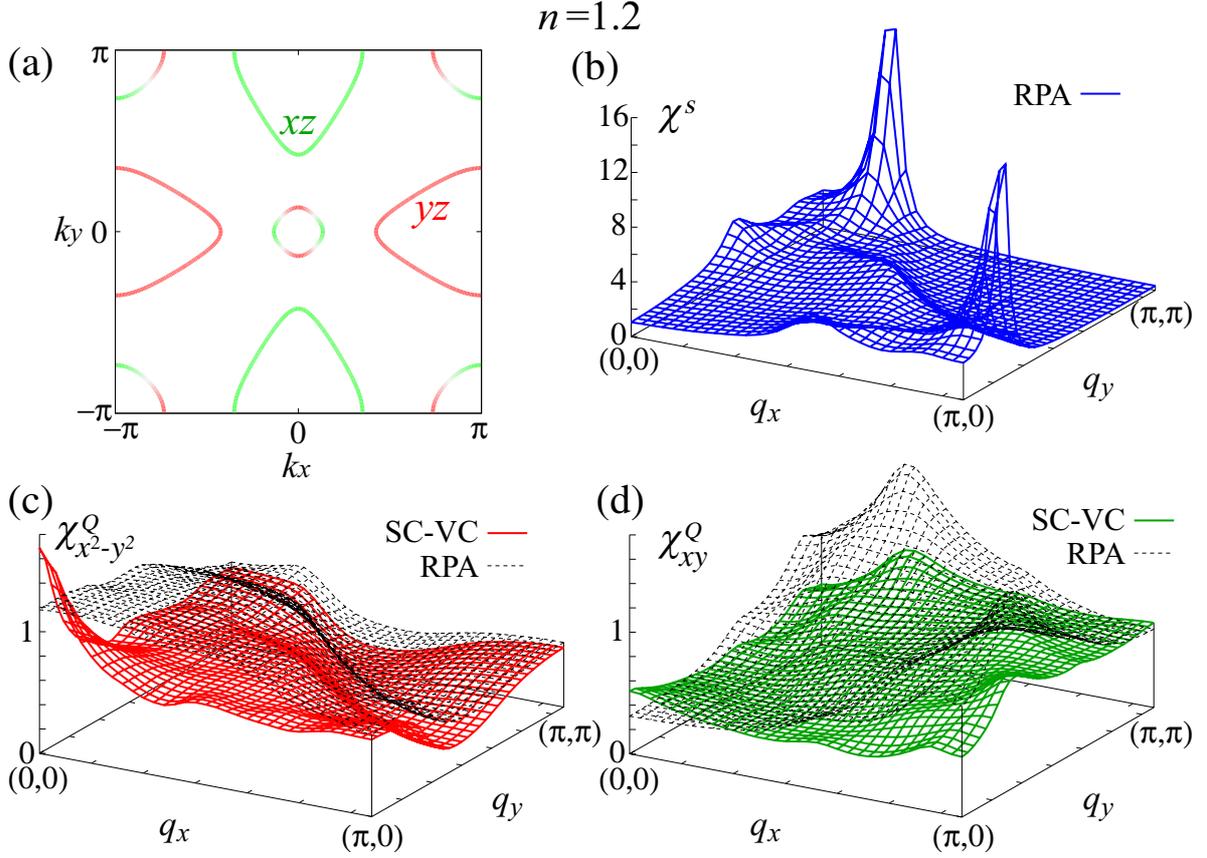}
\caption{(Color online) (a)FS composed of orbital 1 (green line) and orbital
 2 (red line), {\color{black}(b)$\chi^s(\bm{q},0)$ in the RPA,
 (c)$\chi^Q_{x^2-y^2}(\bm{q},0)$ in the SC-VC method (solid red line) and the
 RPA (black dotted line), and (d)$\chi^Q_{xy}(\bm{q},0)$ in the SC-VC
 method (solid green line) and the
 RPA (black dotted line) for $n=1.2$.}
}
\label{n12-case4}
\end{figure}

In Fig. \ref{n07-case1}, we show the results for $n=0.7$, where the
electron FSs disappear.
$\chi^s$ has a small peak around $\bm{Q_4}=(\pi/2,\pi/2)$ due to the weak
nesting between two hole FSs.
In this filling, $\chi^Q_{xy}$ is dominant over $\chi^s$ and 
$\chi^Q_{x^2-y^2}$. The enhancement of $\chi^Q_{x^2-y^2}$ around
$\bm{Q_3}=(\pi,\pi)$ in the SC-VC method is explained by the AL terms  $X^{{\rm
AL},c}_{11,11(22,22)}(\bm{Q_3},0)$. $X^{{\rm
AL},c}_{11,11(22,22)}(\bm{Q_3},0)$ is mainly enlarged by the
term
${V}_{11,11(22,22)}^s(\bm{Q_4},0){V}_{11,11(22,22)}^s(\bm{Q_4},0)$.
}

\begin{figure}
\includegraphics[width=0.7\linewidth]{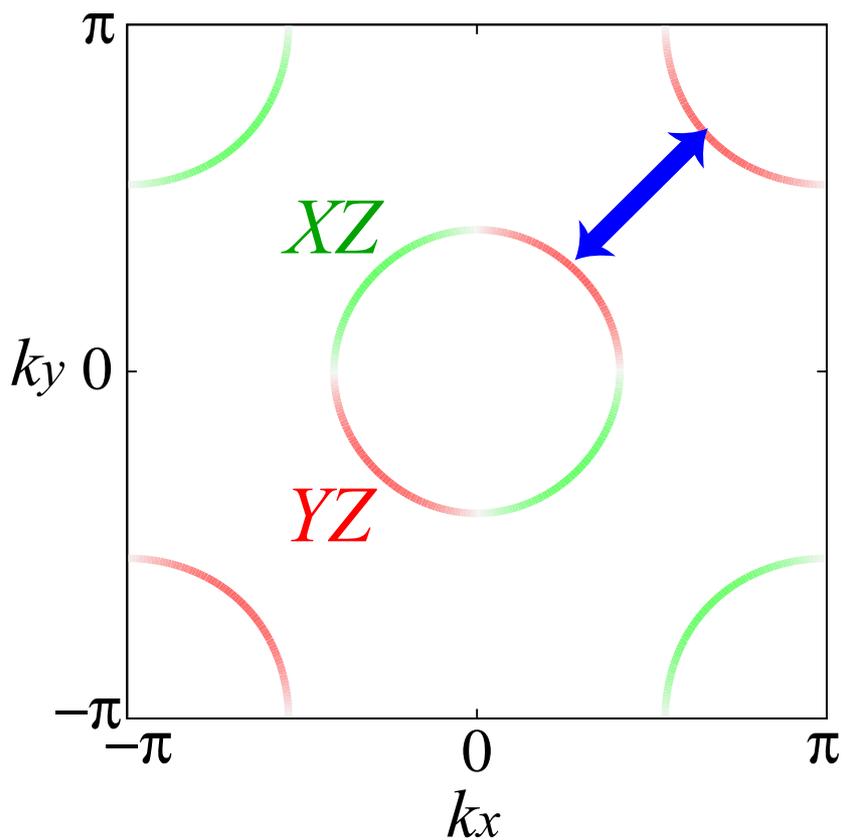}
\caption{(Color online) FSs composed of $d_{XZ}$ orbital (green line) and $d_{YZ}$ orbital
 (red line) for $n=0.7$. The blue two-way arrow denotes the nesting
 vector between the two hole FSs.
}
\label{FS_XY}
\end{figure}

Here, we discuss the reason why the ferro-orbital fluctuation
$\chi^Q_{xy}$ with the peak at $\bm{q}=(0,0)$ is dominant for
$n=0.7$. We introduce the $d_{XZ}$ orbital denoted by the
$45^\circ$-rotated $d_{xz}$
orbital and the $d_{YZ}$ orbital denoted by the $45^\circ$-rotated $d_{yz}$ orbital.
Figure \ref{FS_XY} shows the FSs composed of the $d_{XZ}$ and $d_{YZ}$ orbitals
for $n=0.7$. We see that the nesting vector
$\bm{q}\sim(\pi/2,\pi/2)$ is caused by the intraorbital nesting as
shown by the blue two-way arrow in Fig. \ref{FS_XY}. From
the analogy of the case for $n=1.0$, the intraorbital nesting induces
the ferro-orbital fluctuation $\chi^Q_{X^2-Y^2}$ defined in the
$45^\circ$ rotated basis. Thus, the ferro-orbital fluctuation
$\chi^Q_{xy}$, which is identical to the ferro-orbital fluctuation
$\chi^Q_{X^2-Y^2}$, is dominant for $n=0.7$.
The orbital dependence of FSs is important for determining the dominant orbital
fluctuation.

From these results, we confirmed that dominant fluctuations strongly depend on the
topology and the shape of FSs. {\color{black}We also confirmed that dominant
fluctuations are valid for the small change in the band parameters. Only the
boundary, where dominant fluctuation changes, depends on the band parameters.}

\section{Summary}
We studied the orbital fluctuations near the orbital ordered state using the SC-VC method in the two-orbital
($d_{xz}$, $d_{yz}$) Hubbard model.
We clarified that the nematic order in the non-doped iron-based
superconductors is explained by the two-orbital model around
half filling $n=1.0$. Thus, the $d_{xy}$ orbital is not essential to
realize the nematic order.
Moreover, the $45^\circ$-rotated nematic order
appears in the heavily hole-doped case. 
We confirmed that the dominant orbital fluctuation is determined by the
orbital dependence and the topology of the Fermi surfaces.

\begin{acknowledgment}
We are grateful to H. Kontani for valuable discussions. 
This work was supported by JSPS KAKENHI Grant Number JP17K05543.
Part of numerical calculations was
performed on the Yukawa Institute Computer Facility.
\end{acknowledgment}

%\appendix
%\section{}


\begin{thebibliography}{99}
\bibitem{phase1}
S. Nandi, M. G. Kim, A. Kreyssig, R. M. Fernandes, D. K. Pratt, A. Thaler, N. Ni, S. L. Bud'ko, P. C. Canfield, J. Schmalian, R. J. McQueeney, and A. I. Goldman, Phys. Rev. Lett. {\bf 104}, 057006 (2011). 
\bibitem{phase2}
H. Luetkens, H.-H. Klauss, M. Kraken, F. J. Litterst, T. Dellmann,
	R. Klingeler, C. Hess, R. Khasanov, A. Amato, C. Baines,
	M. Kosmala, O. J. Schumann, M. Braden, J. Hamann-Borrero,
	N. Leps, A. Kondrat, G. Behr, J. Werner, and  B. B\"uchner, Nat. Mater. {\bf 8}, 305 (2009).
\bibitem{Fernandes}
R. M. Fernandes, L. H. VanBebber, S. Bhattacharya, P. Chandra, 
V. Keppens, D. Mandrus, M. A. McGuire, B. C. Sales, A. S. Sefat, 
and J. Schmalian, 
%R.M. Fernandes {\it et al.},
Phys. Rev. Lett. {\bf 105}, 157003 (2010).
\bibitem{Fernandes2}%aniso SDW
R. M. Fernandes, E. Abrahams, and J. Schmalian,
Phys. Rev. Lett. {\bf 107}, 217002 (2011).


\bibitem{Onari-SCVC}
S. Onari and H. Kontani, 
Phys. Rev. Lett. {\bf 109}, 137001 (2012).
\bibitem{Onari-Hdoped}
S. Onari, Y. Yamakawa, and H. Kontani,
Phys. Rev. Lett. {\bf 112}, 187001 (2014).

\bibitem{Kontani-Onari}
H. Kontani and S. Onari, Phys. Rev. Lett. {\bf 104}, 157001 (2010).

\bibitem{Kontani-quad}%aniso SDW
H. Kontani, T. Saito, and S. Onari, Phys. Rev. B {\bf 84}, 024528 (2011).

\bibitem{Kruger}
F. Kr\"{u}ger, S. Kumar, J. Zaanen, and J. van den Brink, Phys. Rev. B {\bf 79}, 054504 (2009).

\bibitem{PP}
%W. Lv, F. Kr\"{u}ger, and P. Phillips, 
%Phys. Rev. B {\bf 82}, 045125 (2010).
 W. Lv, J. Wu, and P. Phillips, 
 Phys. Rev. B {\bf 80}, 224506 (2009).

\bibitem{WKu}
C.-C. Lee, W.-G. Yin, and W. Ku, 
Phys. Rev. Lett. {\bf 103}, 267001 (2009).


%\bibitem{Fang}
%C. Fang, H. Yao, W.-F. Tsai, J. P. Hu, and S. A. Kivelson, Phys. Rev. B
%	{\bf 77}, 224509 (2008).

%\bibitem{Harriger}
%L. W. Harriger, H. Q. Luo, M. S. Liu, C. Frost, J. P. Hu, M. R. Norman,
%	and P. Dai, Phys. Rev. B {\bf 84}, 054544 (2011).




%
%\bibitem{Mazin}
%I. I. Mazin, D. J. Singh, M. D. Johannes, and M. H. Du,
%% I.I. Mazin et al.: 
%Phys. Rev. Lett. {\bf 101}, 057003 (2008).
%
%\bibitem{Tesanovic}
%V. Cvetkovic and Z. Tesanovic, Europhys. Lett. {\bf 85}, 37002 (2009).
%
%\bibitem{Hirschfeld}
%P. J. Hirschfeld, M. M. Korshunov, and I. I. Mazin, Rep. Prog. Phys.
%{\bf 74}, 124508 (2011).
%\bibitem{Chubukov}
%A. V. Chubukov, e-print arXiv:1110.0052.
%\bibitem{Nakai}
%Y. Nakai, T. Iye, S. Kitagawa, K. Ishida, H. Ikeda, S. Kasahara, H. Shishido, T. Shibauchi, Y. Matsuda, T. Terashima
%%Y. Nakai {\it et al.}
%, Phys. Rev. Lett. {\bf 105}, 107003 (2010).
%
%\bibitem{Fujiwara}
%T. Nakano, N. Fujiwara, K. Tatsumi, H. Okada, H. Takahashi, Y. Kamihara, M. Hirano, and H. Hosono,
%%T. Nakano {\it et al.},
%Phys. Rev. B {\bf 81}, 100510(R) (2010).



\bibitem{FeSe-Yamakawa2}
Y. Yamakawa, S. Onari, and H. Kontani, Phys. Rev. X {\bf 6}, 021032 (2016).

\bibitem{FeSe-ARPES62}
Y. Suzuki, T. Shimojima, T. Sonobe, A. Nakamura, M. Sakano, H. Tsuji,
J. Omachi, K. Yoshioka, M. Kuwata-Gonokami, T. Watashige, R. Kobayashi, S. Kasahara,
T. Shibauchi, Y. Matsuda, Y. Yamakawa, H. Kontani, and K. Ishizaka,
Phys. Rev. B {\bf 92}, 205117 (2015).

\bibitem{Onari-FeSe}
S. Onari, Y. Yamakawa, and H. Kontani,
Phys. Rev. Lett. {\bf 116}, 227001 (2016).

\bibitem{Saito}
T. Saito, S. Onari, and H. Kontani, Phys. Rev. B {\bf 82}, 144510
	(2010).
\bibitem{Saito2}
T. Saito, S. Onari, and H. Kontani,
Phys. Rev. B {\bf 83}, 140512(R) (2011).

\bibitem{cuprate1}
Y. Yamakawa and H. Kontani, Phys. Rev. Lett. {\bf 114}, 257001 (2015).

\bibitem{cuprate2}
M. Tsuchiizu, Y. Yamakawa, and H. Kontani, Phys. Rev. B {\bf 93},
	155148 (2016).
\bibitem{Sr327}
M. Tsuchiizu, Y. Ohno, S. Onari, and H. Kontani, Phys. Rev. Lett. {\bf 111}, 057003 (2013).
\bibitem{Sr214-1}
Y. Ohno, M. Tsuchiizu, S. Onari, and H. Kontani, J. Phys. Soc. Jpn. {\bf
	82}, 013707 (2013).
\bibitem{Sr214-2}
M. Tsuchiizu, Y. Yamakawa, S. Onari, Y. Ohno, and H.
Kontani, Phys. Rev. B {\bf 91}, 155103 (2015).
\bibitem{Raghu} S. Raghu, X.-L. Qi, C.-X. Liu, D. J. Scalapino, and
	S.-C. Zhang,
	Phys. Rev. B {\bf 77}, 220503(R) (2008).
{\color{black}
\bibitem{Onari-Springer}
S. Onari and H. Kontani, in {\it Iron-Based Superconductivity},
	ed. P.D. Johnson, G. Xu, and W.-G. Yin, (Springer Verlag Berlin
	and Heidelberg GmbH \& Co. K, 2015) p. 331.
\bibitem{Miyake}
T. Miyake, K. Nakamura, R. Arita, and M. Imada,
J. Phys. Soc. Jpn. {\bf 79}, 044705 (2010).
}
\end{thebibliography}
\end{document}